\documentstyle[aps,epsfig]{revtex}  
\newcommand{\om}{\Omega_{\rm M}}
\newcommand{\ola}{\Omega_{\rm\Lambda}}
\begin{document}
\twocolumn

\title{Photon-axion oscillations and Type Ia supernovae}
\author{Edvard M\"ortsell\thanks{edvard@physto.se}, 
Lars Bergstr\"om\thanks{lbe@physto.se} 
        and Ariel Goobar\thanks{ariel@physto.se}, }
\address{Department of Physics, Stockholm University, \\
         S--106 91 Stockholm, Sweden}
\maketitle
%==========
\section{Abstract}
We compute the probability of photon-axion oscillations in the presence of 
both intergalactic magnetic fields and an electron plasma and investigate 
the effect on
Type Ia supernovae observations. The conversion probability
is calculated using a density matrix formalism by following light-paths through
simulated universes in a Monte-Carlo fashion. We find, that even though the 
effect is highly frequency dependent, one needs to analyze relatively narrow spectral features 
of high redshift objects in order
to discern between the dimming effect from oscillations and a cosmological constant,
in contrast to earlier claims that broad-band photometry is sufficient.

\vspace{3mm}
\noindent
PACS numbers: 13.40.Hq, 98.80.Es, 97.60.Bw

%==========
\section{Introduction}
Recently, Cs\'aki, Kaloper and Terning \cite{csaki} (CKT) proposed that the observed faintness of 
high redshift supernovae (SNe)
could be attributed to the mixing of photons with a light axion in an intergalactic
magnetic field. 
The nature of the oscillations is governed by the strength of the coupling 
which in turn depends on the axion coupling constant mass scale, the photon energy
and the strength of the magnetic field. Assuming magnetic domains
with uncorrelated field direction of size $\sim$ Mpc, CKT found that for optical photons 
the oscillation is maximal and independent of energy, i.e., 
in the limit of infinite travel distance 
one approaches an equilibrium between the two photon polarization states and the axion. 
For optical photons, the probability to detect a photon as a function of the traveled
distance, $l$, was approximated as 
\begin{equation}
  \label{eq:p}
  P_{\rm \gamma}\simeq\frac{2}{3}+\frac{1}{3}\exp{(-l/l_{\rm 0})},
\end{equation}
where $l_{\rm 0}$ is the exponential decay-length. It was claimed that since this effect
will cause additional dimming of high redshift SNe, constraints on the equation
of state parameter of the dark energy component could be significantly relaxed.
(The presence of a non-clustering component of ``dark energy'' has recently
been independently inferred from a combination of measurements of the cosmic
microwave background and the distribution of galaxies on large scales. 
The proposed photon-axion mixing does not remove the need for 
such a component, but its equation of state need not be as close to that 
for a cosmological constant as is the case without such a mixing.)
For low photon energies, the oscillation was found to be energy dependent and the 
mixing very small.
Therefore it should not severely affect the cosmic microwave background radiation
which is redshifted
to low energies at low $z$ where the magnetic fields appear. 
For a magnetic field strength of $|\vec B|\sim 10^{-9}$ G, CKT found that 
the current data can be accommodated by $\om =0.3, \Omega_{\rm X} =0.7, \omega_{\rm X}=-1/3$
if the axion mass is $m\sim 10^{-16}$ eV and the coupling scale is
$M\sim 4\cdot 10^{11}$ GeV.   

It was then pointed out by Deffayet, Harari, Uzan and Zaldarriaga \cite{deffayet} 
that one has to take the effect of 
the intergalactic plasma into account, i.e., the free electrons,
in which the photons are propagating. For a mean electron density of
$n_{\rm e}\sim 10^{-7}\,{\rm cm}^{-3}$, they found that this will alter (lower) 
the oscillation probability 
and may also cause the oscillations for optical photons to be frequency 
dependent 
to a degree where the effect can be ruled out by observational constraints,
namely by studying the color excess between the $B$ and $V$ wavelength bands.     
Only for very definitive properties of the intergalactic magnetic field could
one get a large enough mixing angle and weak enough frequency dependence, 
namely with $|\vec B|\sim 10^{-8}$ 
G over domains of size $\sim 10$ kpc and weaker magnetic field strength over
larger domains. (The strength and spatial properties of intergalactic magnetic are not well constrained by present measurements, for a
recent review see \cite{grasso}.)

In a reply, Cs\'aki et al. \cite{csaki2} pointed out that the mean electron 
density in most of space realistically is lower 
than the estimate used by Deffayet et al.\,by a factor of at least 15. Since the
energy dependence of the oscillations is very sensitive to this value, 
they find that
this factor is enough to bring the energy dependence within current 
experimental bounds.

In this note, we perform a full density matrix calculation of the photon-axion
oscillations and find the conversion probability to be highly frequency 
dependent
and also not necessarily monotonically increasing with increasing redshift. 
We also calculate the color excess between different wavelength bands 
for Type Ia SNe by  
integrating over spectrum templates modified by performing the
density matrix calculation over the appropriate frequency range (taking 
the redshifting of the spectra into account).

%==========
\section{Density matrix formalism}\label{sec:dmf}
We compute the mixing probability using the formalism of density matrices
(see, e.g., \cite{sakurai}).  
Following the notation of \cite{deffayet}, we define the mixing matrix as 
\begin{equation}
  \label{eq:M} 
  M=\left(\begin{array}{ccc}
       \Delta_{\perp} & 0 & \Delta_{\rm M}\cos\alpha\\
       0 & \Delta_{\parallel} & \Delta_{\rm M}\sin\alpha\\
       \Delta_{\rm M}\cos\alpha & \Delta_{\rm M}\sin\alpha & \Delta_{\rm m}
       \end{array}\right).
\end{equation}
The different quantities appearing in this matrix are given by 
\begin{eqnarray}
  \label{eq:terms} 
  \Delta_{\perp} & = & -3.6\times 10^{-25}\left(\frac{\omega}{1\,{\rm eV}}\right)^{-1}
\left(\frac{n_{\rm e}}{10^{-8}\,{\rm cm}^{-3}}\right){\rm cm}^{-1},\\
\Delta_{\parallel} & = & \Delta_{\perp},\\
\Delta_{\rm M} & = & 2\times 10^{-26}\left(\frac{B_{\rm 0,\perp}}{10^{-9}\,{\rm G}}\right)
\left(\frac{M_{\rm a}}{10^{11}\,{\rm GeV}}\right)^{-1}{\rm cm}^{-1},\\
\Delta_{\rm m} & = & -2.5\times 10^{-28}
\left(\frac{m_{\rm a}}{10^{-16}\,{\rm eV}}\right)^2
\left(\frac{\omega}{1\,{\rm eV}}\right)^{-1}{\rm cm}^{-1},
\end{eqnarray}
where $B_{\rm 0,\perp}$ is the strength of the magnetic field perpendicular to the
direction of the photon, $M_{\rm a}$ is the inverse coupling between the 
photon and the axion, $n_{\rm e}$ is the electron density,
$m_{\rm a}$ is the axion mass and $\omega$ is the
energy of the photon.
The angle $\alpha$ is the angle between the (projected) magnetic field
and the (arbitrary, but fixed) perpendicular polarization vector.
Our standard set of parameter-values is given by
\begin{eqnarray}
  \label{eq:values} 
  B_{\rm 0} & = & 10^{-9}(1+z)^2\,{\rm G},\nonumber\\
  M_{\rm a} & = & 10^{11}\,{\rm GeV},\nonumber\\
  m_{\rm a} & = & 10^{-16}\,{\rm eV},\nonumber\\
  n_{\rm e}(z) & = & 10^{-8}(1+z)^3\,{\rm cm}^{-3},
\end{eqnarray}
with a 20 \% dispersion in $B_0$ and $n_{\rm e}$ and the redshift dependence
of the magnetic field strength and the electron density comes from flux 
conservation and cosmological expansion, respectively.
The equation to solve for the evolution of the density matrix $\rho$ is 
given by
\begin{equation}
  \label{eq:rhoeq}
  {\rm i}\delta_{\rm t}\rho =\frac{1}{2\omega}[M,\rho], 
\end{equation}
with initial conditions
\begin{equation}
  \label{eq:rho0} 
  \rho_{\rm 0}=\left(\begin{array}{ccc}
       \frac{1}{2} & 0 & 0\\
       0 & \frac{1}{2} & 0\\
       0 & 0 & 0
       \end{array}\right).
\end{equation}
Here the three diagonal elements refer to two different polarization
intensities and the axion intensity, respectively.
We solve the system of 9 coupled (complex) differential equations 
numerically \cite{lsoda},
by following individual light paths through a large
number of cells where the strength of the magnetic field and the 
electron density is determined from predefined distributions and the 
direction of the magnetic field is random. Through
each cell the background cosmology and the wavelength of the photon 
are updated, as well as the matrices $\rho$ and $M$.

In order to study the qualitative behavior of the solutions, we rewrite
$M$ as a $2\times 2$ matrix,
\begin{equation}
  \label{eq:m2D} 
  M^{\rm 2D}=\left(\begin{array}{ccc}
       \Delta & \Delta_{\rm M}\\
       \Delta_{\rm M} & \Delta_{\rm m}
       \end{array}\right),
\end{equation}
where $\Delta =\Delta_{\perp}=\Delta_{\parallel}$ and 
$\Delta_{\rm M}$ is the component of the magnetic field parallel
to the some average polarization vector of the photon beam.
We solve Eq.~\ref{eq:rhoeq} for the density
matrix $\rho^{\rm 2D}$ with initial conditions
\begin{equation}
  \label{eq:rho02D} 
  \rho_{\rm 0}=\left(\begin{array}{cc}
       1 & 0\\
       0 & 0
       \end{array}\right),
\end{equation}
where the diagonal elements refer to the photon
and the axion intensity respectively. 
Assuming a homogeneous magnetic field and electron density, we can solve the 
two-dimensional system analytically. For the $\rho^{\rm 2D}_{11}$ component,
referring to the photon intensity, we get
\begin{eqnarray}
  \label{eq:rho11} 
  \rho^{\rm 2D}_{11}&=&1-\left(\frac{\Delta_{\rm M}}{2\omega\Omega}\right)^2
        (1-\cos{\Omega t}),\nonumber \\
  \Omega &=&\frac{\sqrt{(\Delta -\Delta_{\rm m})^2+2\Delta_{\rm M}^2}}{2\omega}.
\end{eqnarray}
We see that we get maximal mixing if $\Delta =\Delta_{\rm m}$, i.e., if
$m_{\rm a}=m_{\rm max}
\approx 38\sqrt{n_{\rm e}/(10^{-8}{\rm cm}^{-3})}
\,10^{-16}\,{\rm eV}$. For 
$m_{\rm a}\gg m_{\rm max}$, the 
oscillations are suppressed
as $m_{\rm a}^{-4}$. For 
$m_{\rm a}\ll m_{\rm max}$, the
effect is insensitive to the values of the axion mass. Our numerical simulations
show that even for $m_{\rm a}\approx m_{\rm max}$, results are insensitive to the 
exact value of the axion mass.
For values close to the typical set of parameter-values, we can set 
\begin{equation}
  \label{eq:Omega} 
  \Omega\simeq\frac{\Delta}{2\omega},
\end{equation}
to get
\begin{equation}
  \label{eq:rho11approx} 
  \rho^{\rm 2D}_{11}\simeq 1-\left(\frac{\Delta_{\rm M}}{\Delta}\right)^2
        (1-\cos{\frac{\Delta t}{2\omega}}).
\end{equation}
For small mixing angles, the effect should be roughly
proportional to $B_{\rm 0}^2$ and $M_{\rm a}^{-2}$, whereas the effect should be 
rather insensitive to the exact values of the input parameters in cases of 
close to maximal mixing. 
We also expect the effect to be stronger for low values of the electron
density approaching maximal mixing for $n_{\rm e}=0$. These predictions are confirmed 
by numerical simulations. 
The oscillation length is of the order $\sim$ Mpc. 

%==========
\section{Results}
\begin{figure}[t]
  \centerline{\hbox{\epsfig{figure=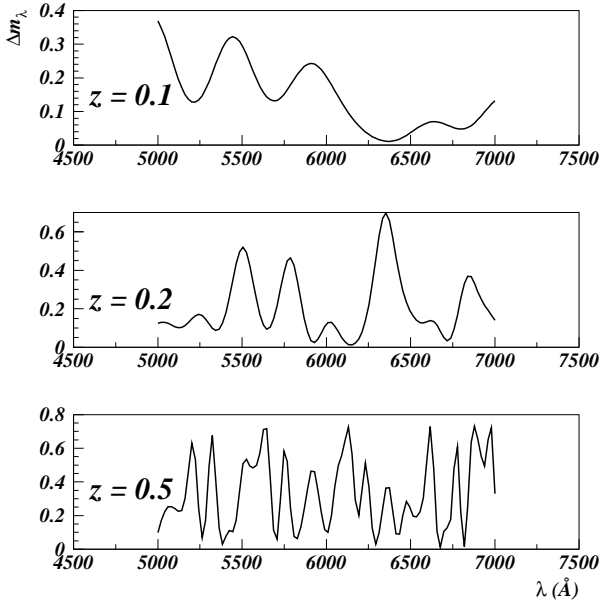,width=0.5\textwidth}}}
  \caption{The attenuation due to photon-axion oscillations for the standard set
        of parameter-values (see  Eq.~\ref{eq:values}) 
        as a function of wavelength at redshift 
        $z=0.1$ (upper panel), $z=0.2$ (middle panel) and $z=0.5$ (lower panel).}
  \label{fig:spectra}
\end{figure}

In Fig.~\ref{fig:spectra}, we show the attenuation due to photon-axion 
oscillations as a function of wavelength for one specific line-of-sight
for our standard set of input parameter-values (see  Eq.~\ref{eq:values}) at 
$z=0.1$ (upper panel), $z=0.2$ (middle panel) and $z=0.5$ (lower panel) 
in a $\om =0.28,\,\ola =0.72$ universe, which is what we will use subsequently.
We have performed a number of simulations using a wide range of cosmological
parameter-values and found the oscillation effect to depend only weakly on
cosmology.
The most dramatic effect is the strong variation of attenuation
with photon energy.

Since the attenuation varies very rapidly with photon energy in a similar
manner over a broad energy range, we expect the frequency dependence to
wash out to large extent when doing broad-band photometry. 

\begin{figure}[t]
  \centerline{\hbox{\epsfig{figure=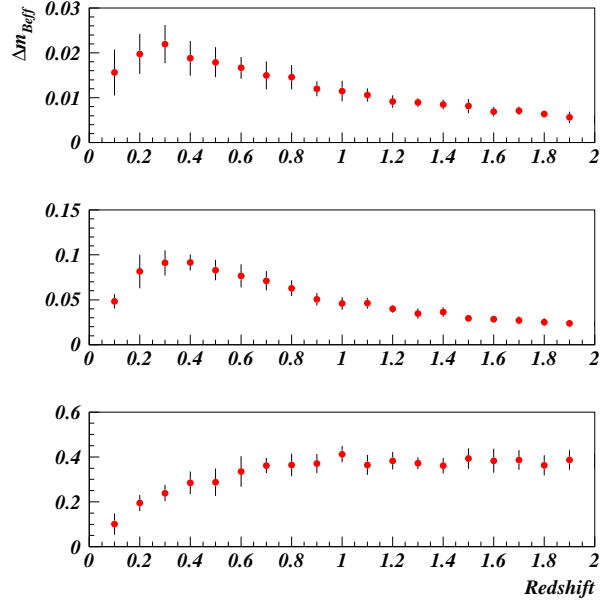,width=0.5\textwidth}}}
  \caption{The attenuation integrated over the rest-frame $B$-band
        of Type Ia SNe due to photon-axion oscillations with 
        $n_{\rm e}=10^{-7}\,{\rm cm}^{-3}(1+z)^3$ (upper panel),
        $n_{\rm e}=5\times 10^{-8}\,{\rm cm}^{-3}(1+z)^3$ (middle panel)
        and $n_{\rm e}=10^{-8}\,{\rm cm}^{-3}(1+z)^3$ (lower panel).
	All other parameter-values are given by Eq.~\ref{eq:values}.}
  \label{fig:n} 
\end{figure}

In Fig.~\ref{fig:n}, we show the rest-frame $B$-band magnitude 
attenuation for Type Ia SNe due to photon-axion 
oscillations for three different values of the electron density,  
in the redshift interval $0<z<2$, using values for the other input parameters
from Eq.~\ref{eq:values}. Each point represents the average value and the
error bars the dispersion for ten different lines-of-sight.
In the upper panel, we have used $n_{\rm e}=10^{-7}\,{\rm cm}^{-3}(1+z)^3$, 
in the middle panel $n_{\rm e}=5\times 10^{-8}\,{\rm cm}^{-3}(1+z)^3$
and in the lower panel $n_{\rm e}=10^{-8}\,{\rm cm}^{-3}(1+z)^3$.
Note that the effect is not necessarily increasing with increasing redshift. 
This is due to the fact that we are studying the {\em rest-frame} $B$-band
magnitude. Since the amplitude of the oscillations scales roughly as 
$(\omega /n_{\rm e})^2$ (see Eq.~\ref{eq:rho11approx}), we need this 
combination to be large at some point in order to get close to 
maximal mixing. If the plasma density is high, the photon energy will
be redshifted to too low energies before the plasma density is diluted 
due to the expansion. 

\begin{figure}[t]
  \centerline{\hbox{\epsfig{figure=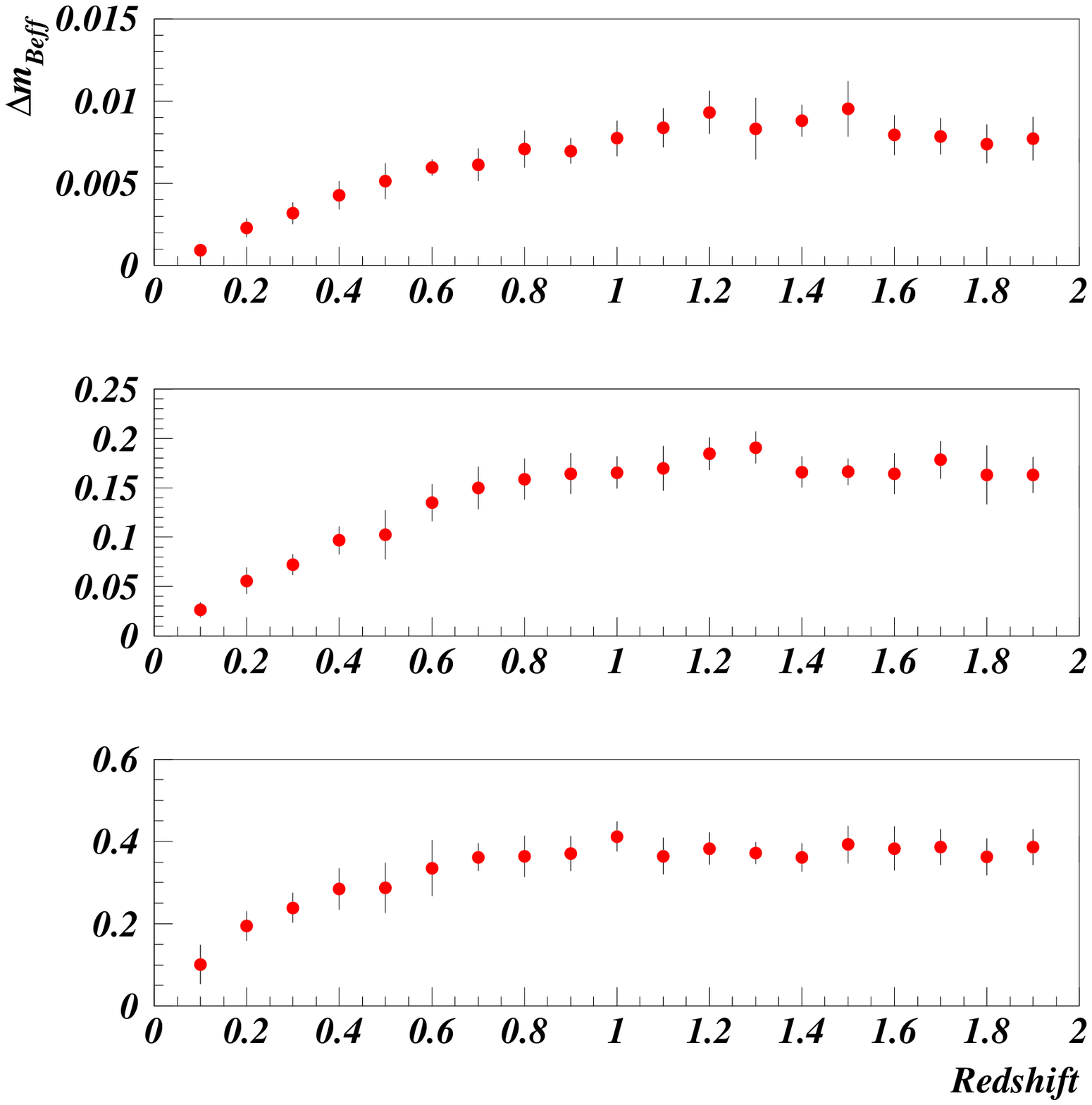,width=0.5\textwidth}}}
  \caption{The attenuation integrated over the rest-frame $B$-band
        of Type Ia SNe due to photon-axion oscillations with 
        $B_{\rm 0}=10^{-10}(1+z)^2\,{\rm G}$ (upper panel),
        $B_{\rm 0}=5\times 10^{-10}(1+z)^2\,{\rm G}$ (middle panel)
        and $B_{\rm 0}=10^{-9}(1+z)^2\,{\rm G}$ (lower panel).
	All other parameter-values are given by Eq.~\ref{eq:values}.}
  \label{fig:b} 
\end{figure}

In Fig.~\ref{fig:b}, the rest-frame $B$-band magnitude 
attenuation for Type Ia SNe for three different values of the
intergalactic field strength is shown.
Again, each point represents the average value and the
error bars the dispersion for ten different lines-of-sight.
In the upper panel, $B_{\rm 0}=10^{-10}(1+z)^2\,{\rm G}$, 
in the middle panel $B_{\rm 0}=5\times 10^{-10}(1+z)^2\,{\rm G}$
and in the lower panel $B_{\rm 0}=10^{-9}(1+z)^2\,{\rm G}$.
All other parameter-values are given by Eq.~\ref{eq:values}.

Our results indicate that we need low electron densities in order to get 
an attenuation that increases with redshift. One should keep in mind that the 
overall normalization can be set by varying the strength of the
magnetic fields and/or the photon-axion coupling strength.

\begin{figure}[t]
  \centerline{\hbox{\epsfig{figure=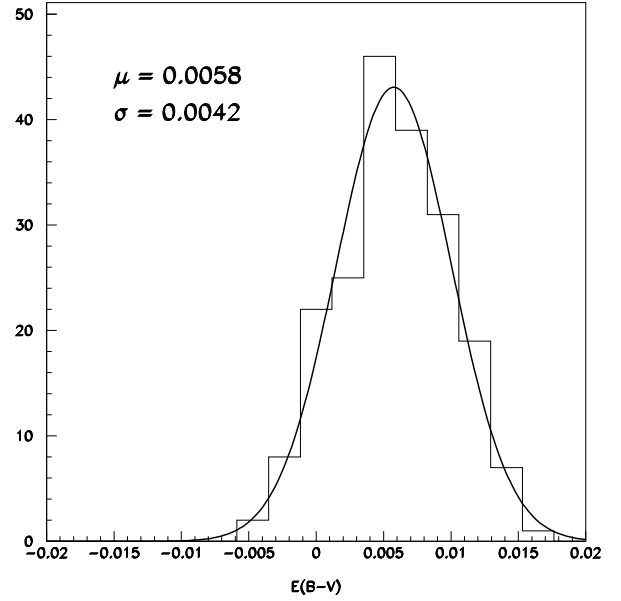,width=0.5\textwidth}}}
  \caption{Color extinctions for rest-frame $E(B-V)$ for 200 Type Ia SNe
       at $z=0.5$.}
  \label{fig:deffayet} 
\end{figure}

Based on a sample of 36 $z\sim0.5$ SNe Perlmutter et al. \cite{perlmutter} measured the
the average rest-frame $B-V$ color excess to be $0.035 \pm 0.022$ mag. We find the expected
color excess with our standard set of parameters to be, 
$E(B-V) = 0.006$, 
with a scatter around this value of 0.004 mag, as shown in Fig.~\ref{fig:deffayet} where
we have simulated 200 Type Ia SNe at z=0.5. 

\begin{figure}[!htb]
  \centerline{\hbox{\epsfig{figure=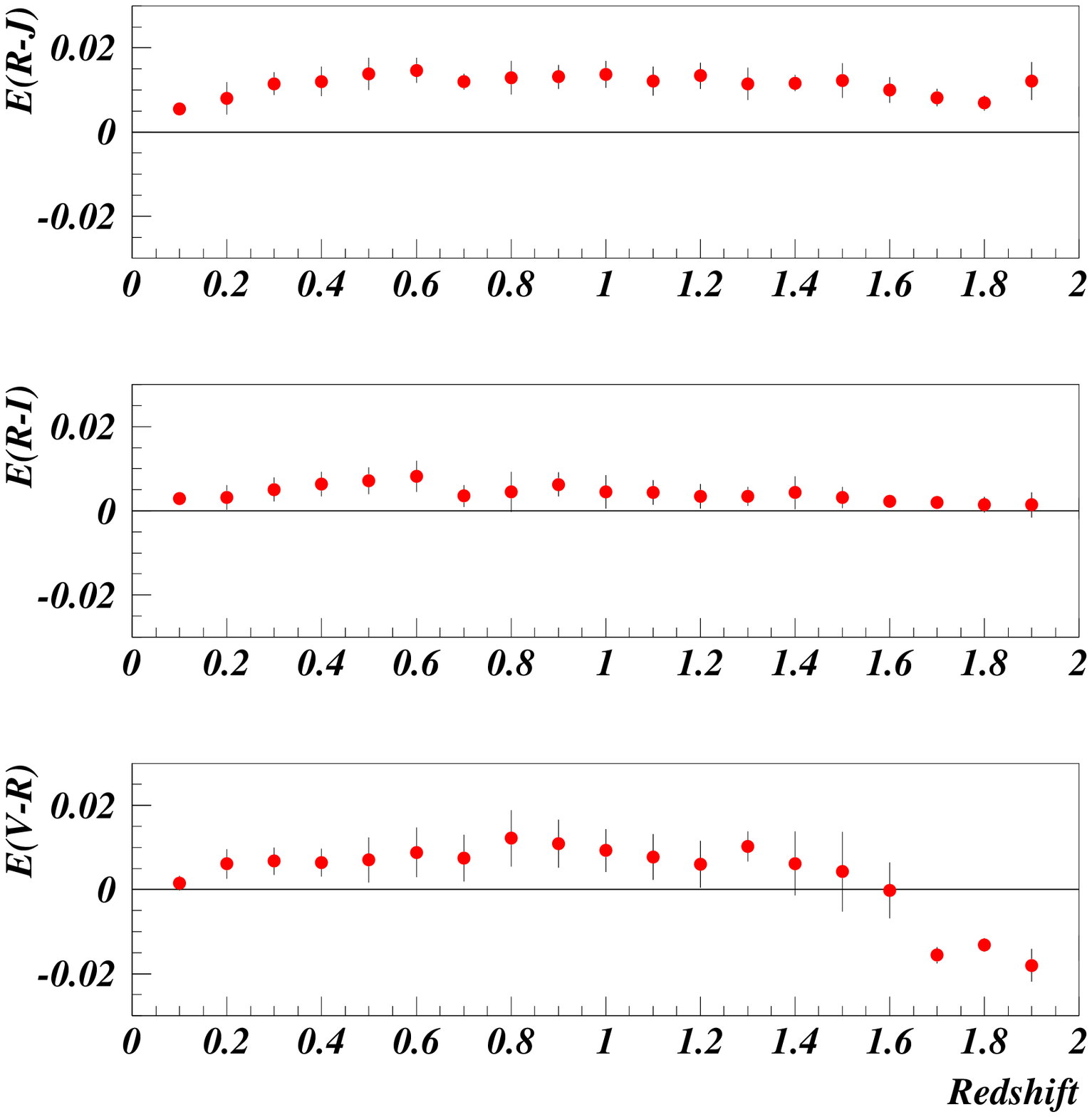,width=0.5\textwidth}}}
  \caption{Color extinctions $E(V-J)$, $E(R-J)$ and $E(I-J)$ for Type Ia SNe
        as a function of redshift. Note that the dimming associated with photon-axion mixing, 
        unlike extinction by dust, also can generate {\em blueing}.}
  \label{fig:mixcolors} 
\end{figure}

We can investigate the frequency dependence when doing broad-band photometry
by studying the  extinction (at maximum intensity) in $V-J$, $R-J$ and $I-J$ for 
Type Ia SNe as a function of redshift, as we show in Fig.~\ref{fig:mixcolors}.
All the broad-band filters and spectroscopy wavelength scales are in the observer's
frame. We can see that in all three cases, the color excess is very small 
and thus difficult to measure. 

In general, the nature of the frequency dependence will depend on the exact values 
of the input parameters with the possibility of generating both reddening and 
blueing, where in some cases the nature of the effect will depend on redshift
(see, e.g., lower panel of Fig.~\ref{fig:mixcolors}).  
We thus conclude that it will be very difficult to use the color excess
between different broad-bands to put severe limits on the photon-axion mixing parameters. 

%==========
\section{Discussion}

Our numerical simulations indicate that in order to get a dimming
effect from photon-axion oscillations similar to the one from a
cosmological constant (increasing at lower redshifts, saturating at
higher), one would need to have an average intergalactic electron
density of $n_{\rm e}\lesssim 10^{-8}\,{\rm cm}^{-3}(1+z)^3$. 
Assuming this, it should be possible to vary the average magnetic
field strength or the photon-axion coupling strength in
order to fit the current broad-band photometry data.
Note that in the case of close to maximal mixing, results are generally 
not very sensitive to the exact values of the input parameters, yielding
results similar to the upper panel in  Fig.~\ref{fig:n}.

Depending on the choice of mixing parameters
the color excess terms of sources at cosmological distances 
could be either positive or negative, the latter
case being particularly interesting as it could not be confused with regular
extinction by dust. Since photon-axion oscillations can cause either
reddening or blueing (or no color excess at all) for close to maximal 
mixing and the integrated broad-band magnitudes wash out the dispersion in attenuation, 
we expect spectroscopic studies of high-z objects to be a more powerful discriminator between
different oscillation models and the case of a cosmological constant
or any other dark energy component (for which we suppose the dimming to be
entirely frequency independent). 
Systematic analysis of quasar, gamma-ray burst and galaxy spectra as 
a function of redshift by, e.g., the Sloan Digital Sky Survey and 2dF 
groups are probably the best probes
for the photon-axion mixing parameter space. Note that the source size sets
a lower limit to the size of the fluctuations that can be probed since the 
fluctuations will average out if photons from different 
parts of the source will
travel through different magnetic field strengths and electron densities.

SN spectra, available with good
signal-to-noise ratios up to at least $z\sim 0.5$, are 
useful probes for the large
mixing parameters that would yield a sufficient dimming of Type Ia SNe as to 
explain the Hubble diagrams in \cite{perlmutter,riess} 
without invoking dark energy.

\section*{Acknowledgements}
We are grateful to S.~Hansen for bringing the CKT paper to our
attention, and to G.~Raffelt for insightful comments improving the 
quality of the manuscript. We also wish to thank 
J.~Edsj\"o, H.~Rubinstein, C.~Fransson and J.~Silk
for helpful discussions.
The research of L.B.~is sponsored by the Swedish Research Council (VR).
A.G. is a Royal Swedish Academy 
Research Fellow supported by a 
grant from the Knut and Alice Wallenberg Foundation.

%========== 
\end{document}